\documentclass{article}
\usepackage{spconf,amsmath,graphicx}
\usepackage{comment}
\usepackage{subfig}
\usepackage{colortbl}
\usepackage{soul}
\usepackage{cite}

\usepackage{adjustbox}
\usepackage{tabularx}
\usepackage{multirow}
\usepackage{threeparttable}

\usepackage{amsfonts}

\usepackage{amsmath}

\title{RadarEye: Robust Liquid Level Tracking using mmWave Radar in Robotic Pouring}
\name{Hongyu Deng and He Chen
}
\address{Department of Information Engineering, The Chinese University of Hong Kong, Hong Kong SAR, China}

\begin{document}

\maketitle

\begin{abstract}
Transparent liquid manipulation in robotic pouring remains challenging for perception systems: specular/refraction effects and lighting variability degrade visual cues, undermining reliable level estimation. To address this challenge, we introduce RadarEye, a real-time mmWave radar signal processing pipeline for robust liquid level estimation and tracking during the whole pouring process. RadarEye integrates (i) a high-resolution range–angle beamforming module for liquid level sensing and (ii) a physics-informed mid-pour tracker that suppresses multipath to maintain lock on the liquid surface despite stream-induced clutter and source container reflections. The pipeline delivers sub-millisecond latency. In real-robot water-pouring experiments, RadarEye achieves a 0.35 cm median absolute height error at 0.62 ms per update, substantially outperforming vision and ultrasound baselines.
\end{abstract}

\section{Introduction}
\begin{figure}[]
    \centering
    \subfloat[Depth errors in transparent liquid level estimation.]{
        \includegraphics[width = 0.4\textwidth]{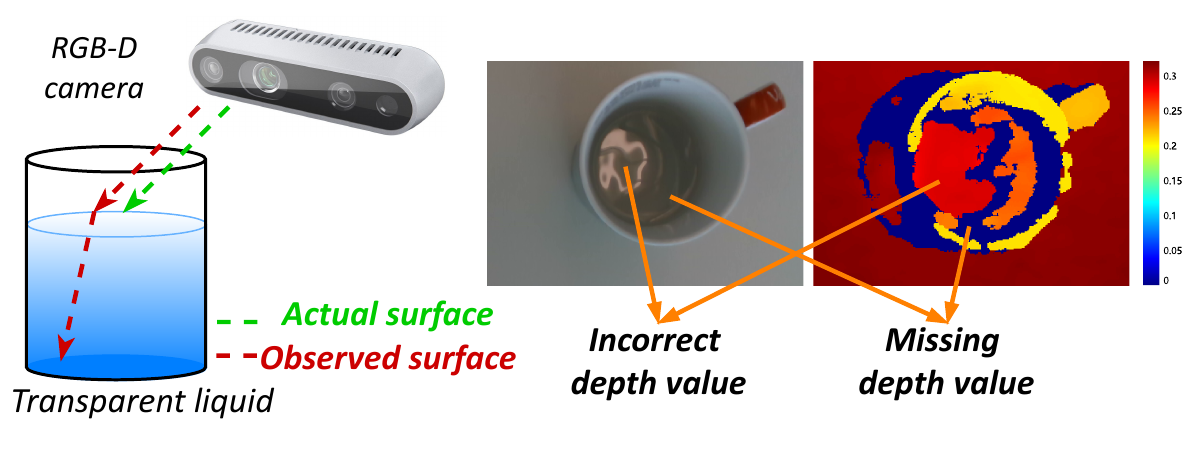} 
    }
    \\
    \subfloat[RadarEye robustly tracks liquid levels using mmWave radar.]{
        \includegraphics[width = 0.4\textwidth]{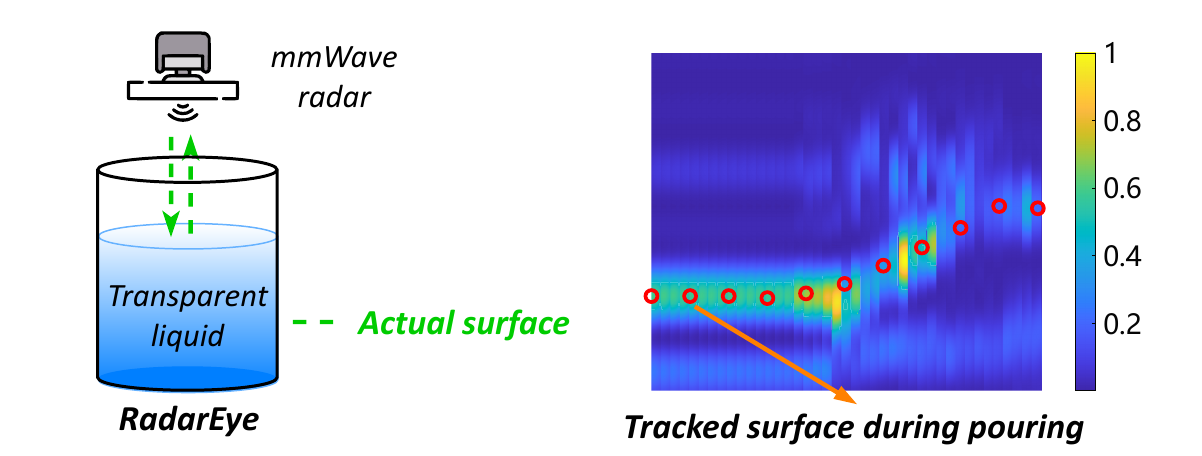} 
    }
    \caption{Conventional vision struggles with transparent liquids (top), while RadarEye leverages mmWave radar to overcome this challenge (bottom).
}    
    \label{motivation_icassp}
    \vspace{-1.5em}
\end{figure}

Pouring transparent liquids like water remains challenging for robots. Despite advances in computer vision, the refraction and specular reflection of transparent liquids violate the geometric assumptions underlying standard vision pipelines. As a result, vision-based liquid level estimation is often unreliable. Moreover, optical cameras are highly sensitive to  illumination changes, further limiting the reliability of conventional vision-based detection. To illustrate these issues, we ran a preliminary study with an RGB-D camera to estimate transparent liquid levels. As shown in Fig. \ref{motivation_icassp} (a), two dominant failure modes emerge: (1) missing depth at the liquid surface due to specular reflections, and (2) biased depth caused by geometric distortion when sensing through the liquid. These observations align with prior reports \cite{cleargrasp, deng2025fusegrasp, do2016probabilistic}.

Despite recent progress, the intrinsic optics of transparent liquids still make robotic pouring difficult. Prior work has explored vision-based level estimation using probabilistic models and neural networks \cite{do2016probabilistic, mottaghi2017see, lin2023pourit}. However, these vision-centric methods are fragile to lighting variability, and neural models often incur inference latencies that hinder true real-time tracking. To address these limitations, researchers have turned to alternative sensor modalities, such as acoustic, capacitive, and tactile sensing, that leverage the physical properties of liquids for more reliable perception \cite{bagad2025sound, wang2023capacitive, xompero2022audio, xu2024robot, li2022see, piacenza2022pouring}. 
Nevertheless, these approaches infer liquid level indirectly from secondary cues, such as flow acoustics or container vibrations, rather than measuring the liquid surface directly. Moreover, their performance often hinges on controlled lab conditions, limiting real-world applicability, and they can incur high response latency. This motivates a new sensing modality that enables direct, robust, and real-time liquid-level tracking.

To address the perception challenges in transparent-liquid pouring, we propose a mmWave radar-based method that directly measures liquid level in real time. mmWave radar is a promising sensing modality for this task. Its high carrier frequency yields fine spatial resolution and robust perception in challenging conditions \cite{zheng2021more, zhang2023rf}. The short wavelengths promote strong scattering, which is advantageous for detecting transparent media. In particular, the reduced penetration at mmWave frequencies makes transparent liquids effectively opaque, enabling reliable detection of the liquid level~\cite{smooth_reflect1}. Consequently, a pouring robot equipped with mmWave radar can generalize across diverse containers while achieving precise, real-time liquid-level tracking. {However, building a robust, real-time signal-processing pipeline for radar-based liquid-level tracking is nontrivial. Radar returns are sparse and often contaminated by multipath. While beamforming can detect reflections from a static liquid surface, its effectiveness degrades during continuous pouring. As the liquid level rises, surface reflections become increasingly hard to distinguish from concurrent echoes off the robot gripper, workspace fixtures, and the source and target containers. This ambiguity hinders continuous tracking and calls for an algorithm that can reliably isolate and follow the true liquid-level signal amid interfering reflections.
}

{To address this challenge}, we introduce RadarEye, a robust, real-time mmWave radar pipeline for liquid-level tracking in robotic pouring, as illustrated in Fig. \ref{motivation_icassp} (b). RadarEye uses a mmWave radar to synthesize a virtual antenna array and applies spatial signal processing to the received echoes to produce a high-resolution 2D spectrum over angle of arrival (AoA) and time of flight (ToF) for estimating liquid-surface reflection. This spectrum supports direct spatial localization and range estimation, markedly improving liquid-level perception. To resolve ambiguities during dynamic pouring, RadarEye incorporates a physics-informed optimization that suppresses multipath interference, enabling fast, continuous tracking of the liquid level and thereby improving overall pouring accuracy. In our real-world robotic experiments of water pouring, RadarEye achieves a median level-estimation error of 0.35 cm with 0.62 ms latency, substantially outperforming all evaluated baselines.

\section{System Design}\label{sec_design}
\subsection{{Radar Beamforming for Liquid Level Estimation}}
RadarEye positions the mmWave radar directly above the liquid container to maximize reflections from the liquid surface. The reflected signals from environmental objects, including the liquid surface, target and source containers, and robot gripper, are captured by a uniform linear array. The received signal can be written as
\begin{equation}\label{eq_rece}
		r_{m,k}(t) = \sum_{l=1}^{L}s(t)  \alpha_{m,l}(t) e^{-j 2\pi f_k\tau_l(t)} e^{-j 2\pi f_k \frac{(m-1)d\cos\theta_l(t)}{c}},
\end{equation}
where $m$, $k$, $l$, and $t$ denote the receiver antenna index, frequency-domain sampling point index, propagation path index, and time slot, respectively. $L$ is the total number of propagation paths, $s$ is the transmitted signal, $\alpha_{m,l}$ is the complex attenuation coefficient, $f_k$ denotes the signal frequency, $d$ is the array spacing, and $c$ is the speed of light. $\theta_l$ and $\tau_l$ represent the AoA and ToF of the $l$-th path, respectively. The received signals from all antennas and frequency points are concatenated into a vector ${\bf{r}}(t)$, defined as $ {\bf{r}}(t) = [r_{1,1}(t), \cdots, r_{1,K}(t), \cdots, r_{M,1}(t) \cdots, r_{M,K}(t)]^T$, where $M$ and $K$ denote the number of receiver antennas and frequency points, respectively. 

We detect the liquid level by inferring all feasible propagation paths from the received signals. Given that $\theta$ and $\tau$ are continuous, we probe the AoA–ToF plane associated with the linear array via beamspace steering and delay focusing to enumerate candidate paths. Specifically, we discretize the AoA-ToF plane into an $N \times N$ grid and index the bins as $(i,j)$ with $i,j \in {1,\ldots,N}$. To detect the signal strength of the $(i,j)$-th bin, we define the following steering vector
\begin{equation}\label{eq_phase-shift-vector}
{\bf{a}}_{i,j} = {[\phi_{1,1}(i,j), \cdots, \phi_{m,k}(i, j), \cdots, \phi_{M,K}(i,j)]^T},
\end{equation}
where $\phi_{m,k}(i,j) = e^{-j2\pi f_k\tau_j}   e^ {-j\frac{2\pi f_k(m-1)d \cos \theta_i}{c}}$.

\begin{figure}[]
    \centering
    \subfloat[{Static liquid surface.}]{
        \includegraphics[width = 0.23\textwidth]{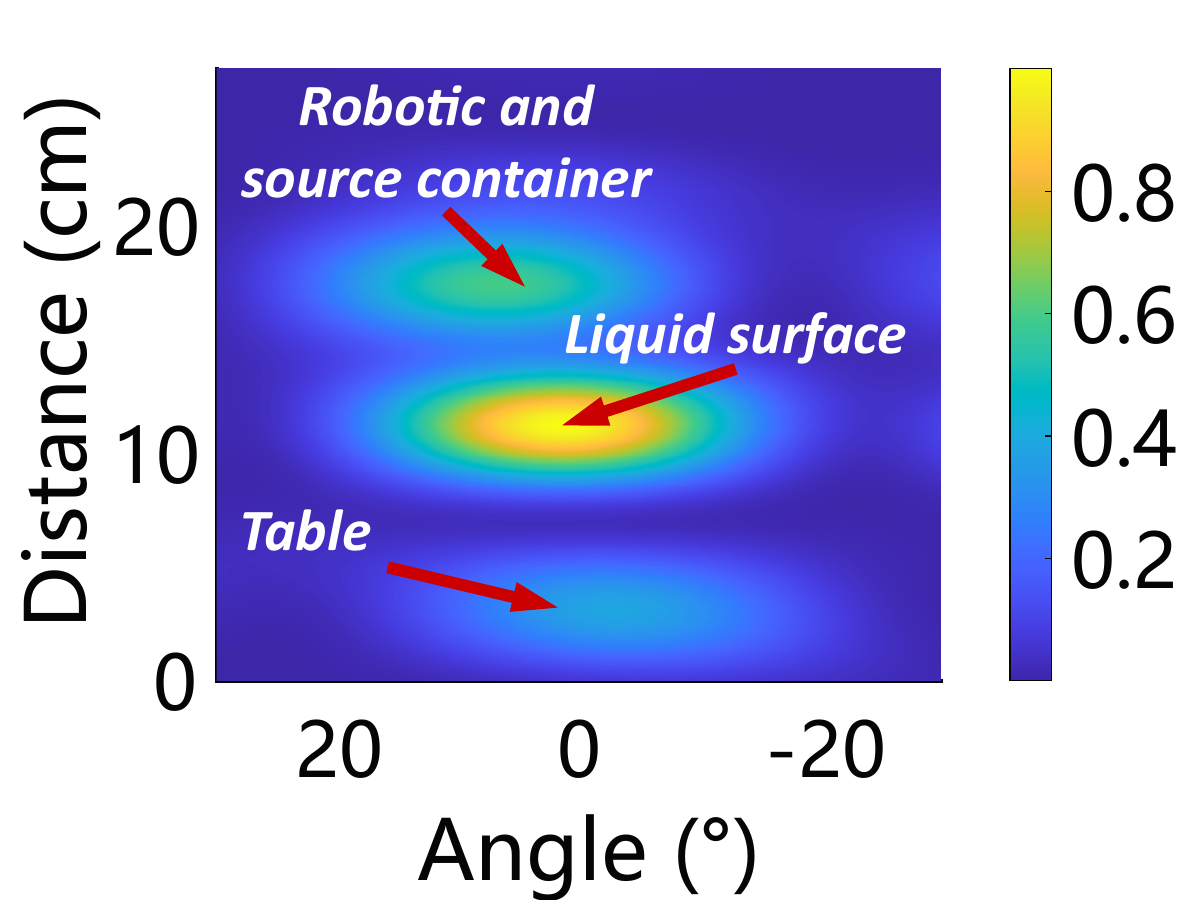} 
    }
    \subfloat[{Dynamic liquid surface.}]{
        \includegraphics[width = 0.23\textwidth]{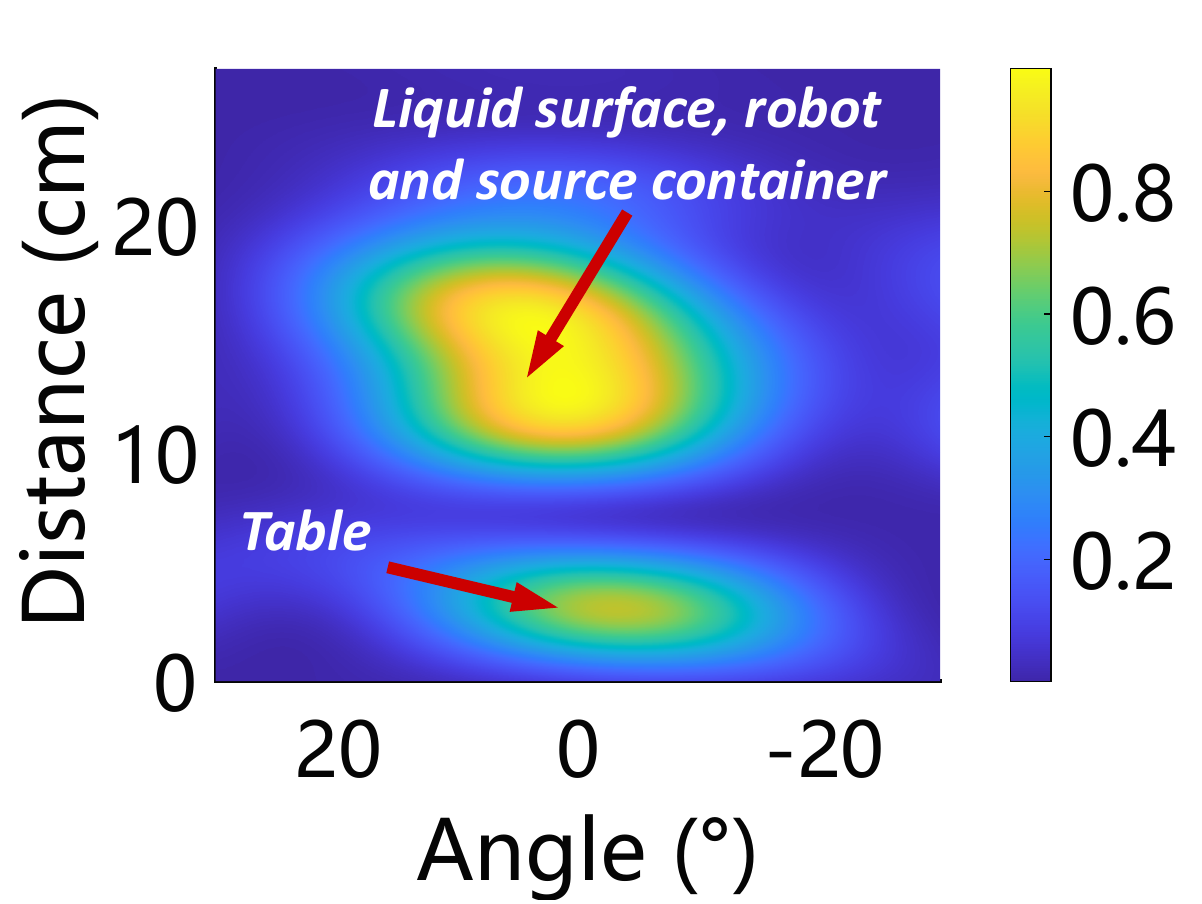}  
    }
    \caption{AoA–ToF spectra for static vs. dynamic (pouring) water surfaces.}    
    \label{fig_challenges}
    \vspace{-1.5em}
\end{figure}

\begin{figure*}[]
\centering
\begin{minipage}{0.56\textwidth}
\centering
\subfloat[AoA-ToF spectrum evolution during pouring.]{
    \includegraphics[width = 0.325\textwidth]{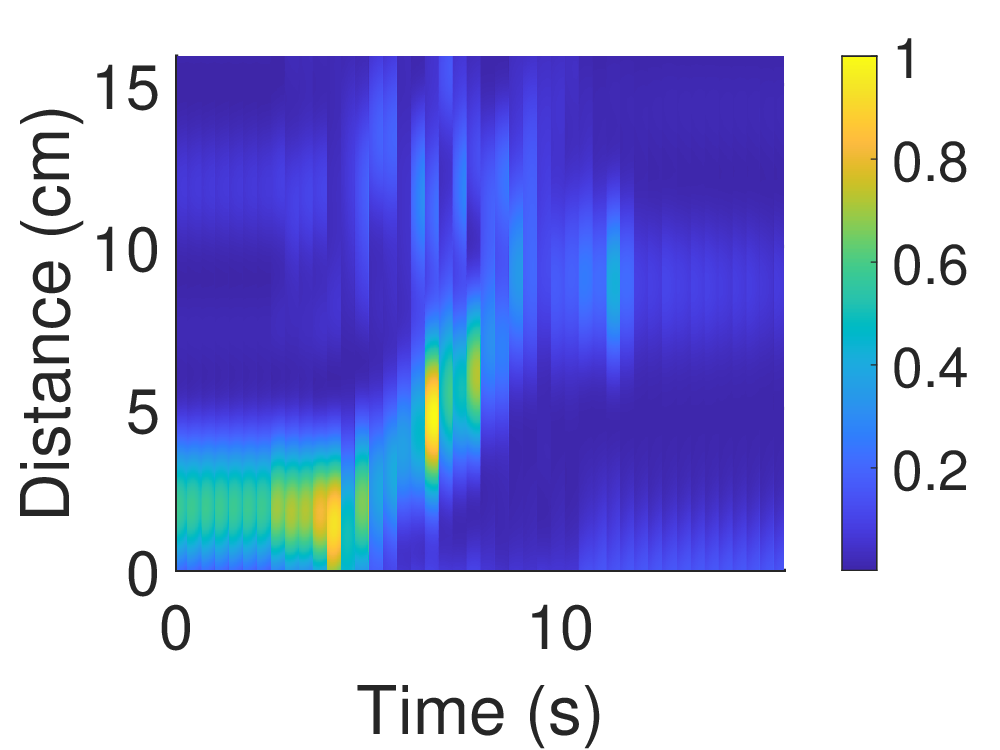} 
}
\subfloat[Peak detection algorithm.]{
    \includegraphics[width = 0.325\textwidth]{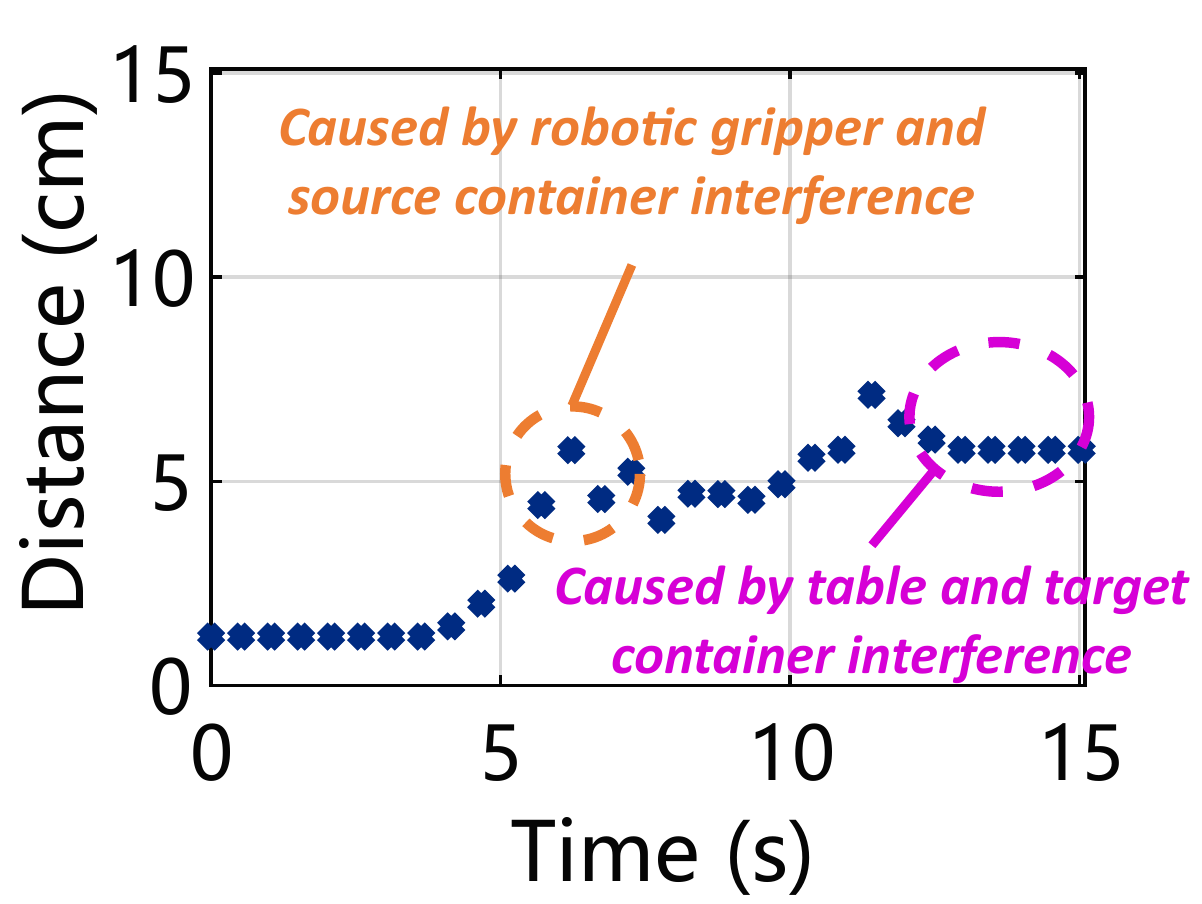} 
}
\subfloat[Our physics-informed algorithm.]{
    \includegraphics[width = 0.325\textwidth]{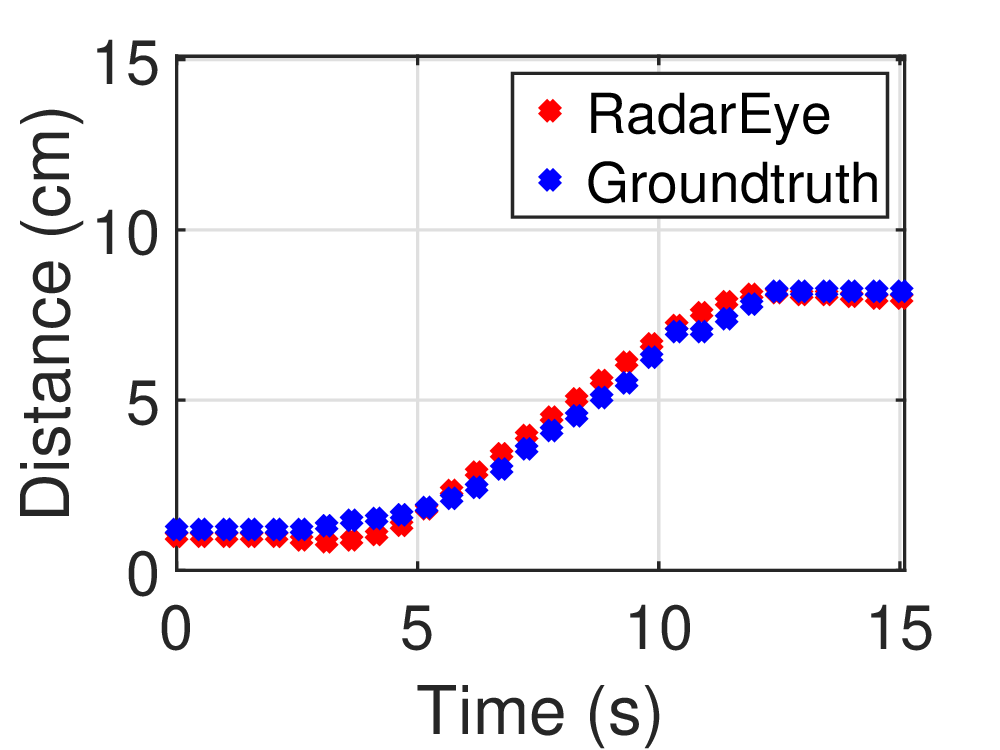} 
}
\caption{Liquid surface tracking during water pouring.}   
\label{fig_design_express}
\end{minipage}
\hfill
\begin{minipage}{0.42\textwidth}
\centering
\includegraphics[width=1\linewidth]{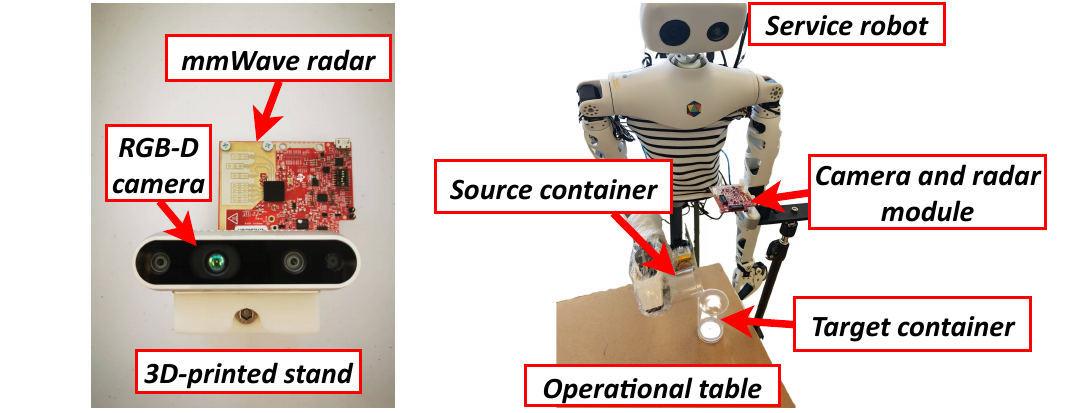} 
\caption{Experiment implementation.}
\label{experiment}
\end{minipage}
\vspace{-1.5em}
\end{figure*} 

{The signal strength associated with the $(i, j)$-th bin can be obtained by $p_{i,j} = {|\bf{a}}_{i,j}^H  {\bf{r}}|$. The process first compensates for phase shifts across adjacent antennas and frequency sampling points, then performs coherent summation over all antennas and frequencies. Signals originating from the $(i,j)$ bin add coherently, whereas others average out. We define the AoA–ToF spectrum $\mathbf{P}$ with entries $\mathbf{P}(i,j) = p_{i,j}$. With a sufficiently dense AoA–ToF grid, the spectrum ${\bf P}$ attains high values at grid points aligned with the true reflected paths.

In measurements of static liquid surfaces, the surface serves as the dominant reflector, producing the strongest return because the radar is located above the target container and the gripper/source container lies to the side. Consequently, the AoA–ToF bin index for the surface path, $(i^*, j^*)$, achieves the highest magnitude in $\mathbf{P}$. That is,
\begin{equation}\label{eq_opt}
(i^*,j^*) = \arg\max_{(i,j)}{\bf{P}}({i,j}).
\end{equation}
Note that in (\ref{eq_opt}), we assume that the AoA-ToF bin index of the line-of-sight path was not included in $\bf P$. In this case, echoes from the liquid surface are clearly resolved under static conditions, as shown in Fig.~\ref{fig_challenges}~(a).

\vspace{-1em}
\subsection{Physics-Informed Liquid Level Tracking}
In pouring scenarios, the radar captures a composite of reflections: the liquid surface, the source container/robot gripper, and the operational desktop, which occupy the AoA–ToF regions above and below the liquid surface. These components overlap in the AoA–ToF domain, confounding liquid-level tracking (see Fig.~\ref{fig_challenges}(b) as a typical example). Static clutter (e.g., the desktop) can be mitigated via differential measurements, but dynamic returns from the liquid, container, and gripper persist as aliased interference, markedly reducing accuracy. Fig.~\ref{fig_design_express}~(a) demonstrated an AoA-ToF spectrum evolution  during an robotic water pouring task. It can be seen that during the period from 4-12s, two distinct reflection lines are formed in the AoA-ToF spectrum. The lower path corresponds to the actual liquid surface height, while the upper path is caused by reflections from the robotic gripper and the source container. Such interference can easily cause RadarEye to misestimate the liquid surface.

Conventional approaches, such as smoothing filters and peak detection algorithms, often fail to reliably address this issue. This failure is primarily due to scenarios where the magnitude of interfering reflections exceeds that of the true echo from the liquid surface. Consequently, directly selecting the bin with the highest spectral magnitude frequently leads to significant disturbances in the estimated liquid level curve. This phenomenon is illustrated in Fig. \ref{fig_design_express}(b). An alternative approach involves training a lightweight deep learning model to estimate the liquid level directly from the AoA-ToF spectrum. However, this method typically requires a large volume of training data and often generalizes poorly to unseen scenarios. Further, such models can introduce non-negligible inference latency. Given the real-time requirements of liquid pouring, these delays can impair system performance.

To enable accurate, real-time liquid-level tracking, RadarEye uses a physics-informed tracker that leverages the slow temporal evolution of liquid dynamics. Rather than selecting peak magnitudes from the AoA–ToF spectrum $\bf{P}$ at each timestamp, we jointly estimate the continuous liquid-surface indices $(i, j)$'s through temporal optimization. To that end, we define ${\bf P}_t$ as the AoA-ToF spectrum at slot $t$. We then define the cost function of the transition from the $(i,j)$-th bin in ${\bf P}_t$ to the $({i'},{j'})$-th bin in ${\bf P}_{t+1}$ as
\begin{equation}\label{eq_costfunc}
\begin{aligned}
c_t\left((i,j)\to({i'},{j'})\right) =& -{\bf{P}}_t{(i,j)} - {\bf{P}}_{t+1}{(i',j')} + \\
&\omega \eta\left((i,j)\to({i'},{j'})\right),
\end{aligned}
\end{equation}
where $\eta\left((i,j)\to({i'},{j'})\right) = \omega_\theta |i - i'|_2 + \omega_\tau |j - j'|_2$ quantifies the displacement between two bins in the consecutive slots. Weighting factors $\omega_\theta$ and $\omega_\tau$ are empirically set to 0.1.  Since the liquid surface evolves continuously, valid transitions produce small values of the $\eta$ function, whereas large values typically indicate discontinuous multipath reflections or other non-liquid signals. 

Based on prior experiments, we assume that the initial estimation of the static liquid level is reliable. Once pouring begins, RadarEye needs to estimate the liquid level at the end of each slot in real time. The total number of possible bin transition paths from the starting point to the current slot $T$ is $N^{2(T-1)}$. We formulate liquid level tracking as an optimal path search over a graph spanning all time steps till $T$, where nodes represent candidate AoA-ToF bins and edge weights correspond to a transition cost function $c$. The optimal path must satisfy two constraints: (1) sufficient signal strength at each node, and (2) smooth transitions between consecutive nodes. This leads to the following optimization problem solved at the end of each slot:
\begin{equation}\label{eq_opt_trajectory}
\delta^* = \arg\min_{\delta \in \Delta_T} C(\delta), \; \text{where} \; C(\delta) = \sum_{t=0}^{T-1} c_t(\delta(t)),
\end{equation}
where $\Delta_T$ denotes the set of all possible transition paths up to slot $T$, and $\delta(t)$ represents the bin transition from slot $t$ to $t+1$ in a given path $\delta$. The most recent liquid level estimate corresponds to the final bin in the optimal path $\delta^*$. To reduce computational complexity, we exploit the smooth nature of liquid level dynamics by constraining each transition to a $Q^2$-neighborhood around the current bin ($Q \ll N$). As shown in Fig.~\ref{fig_design_express}(c), our method reliably tracks the dynamic liquid level.

\begin{figure*}[]
\centering
\begin{minipage}{0.42\textwidth}
\centering
\includegraphics[width=1\linewidth]{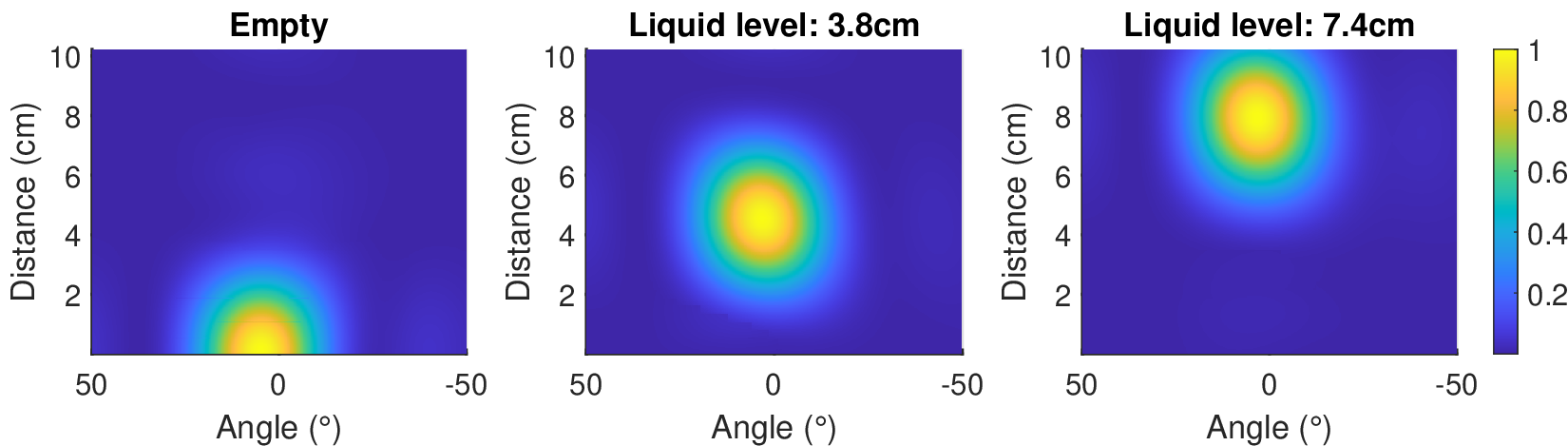} 
\caption{Water level estimation under incremental
filling.}	
\label{radar_performance}
\end{minipage}
\hfill
\begin{minipage}{0.56\textwidth}
\centering
\subfloat[Pouring curve.]{
    \includegraphics[width = 0.325\textwidth]{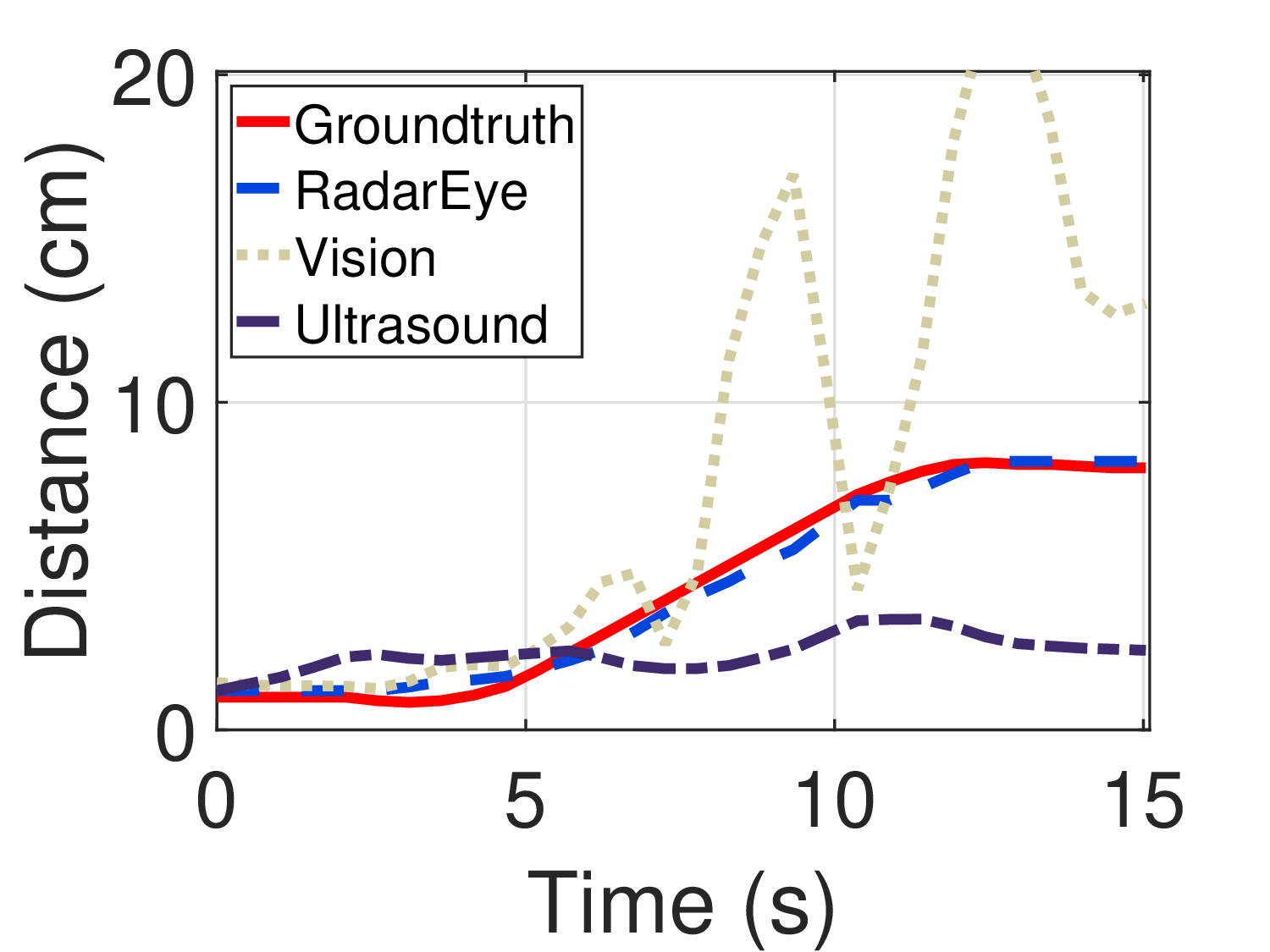} 
}
\subfloat[Level estimation errors.]{
    \includegraphics[width = 0.325\textwidth]{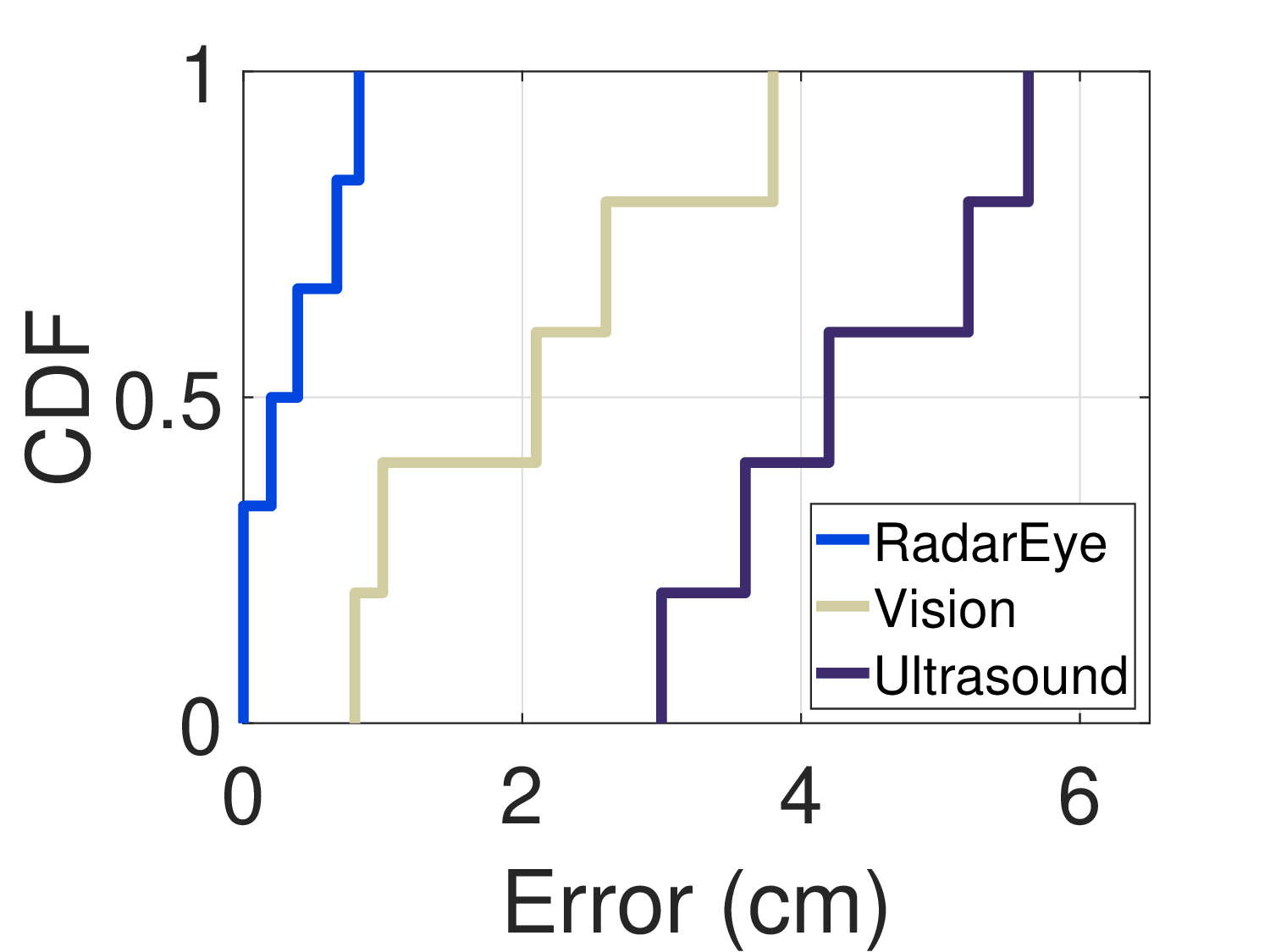} 
}
\subfloat[Response time.]{
    \includegraphics[width = 0.325\textwidth]{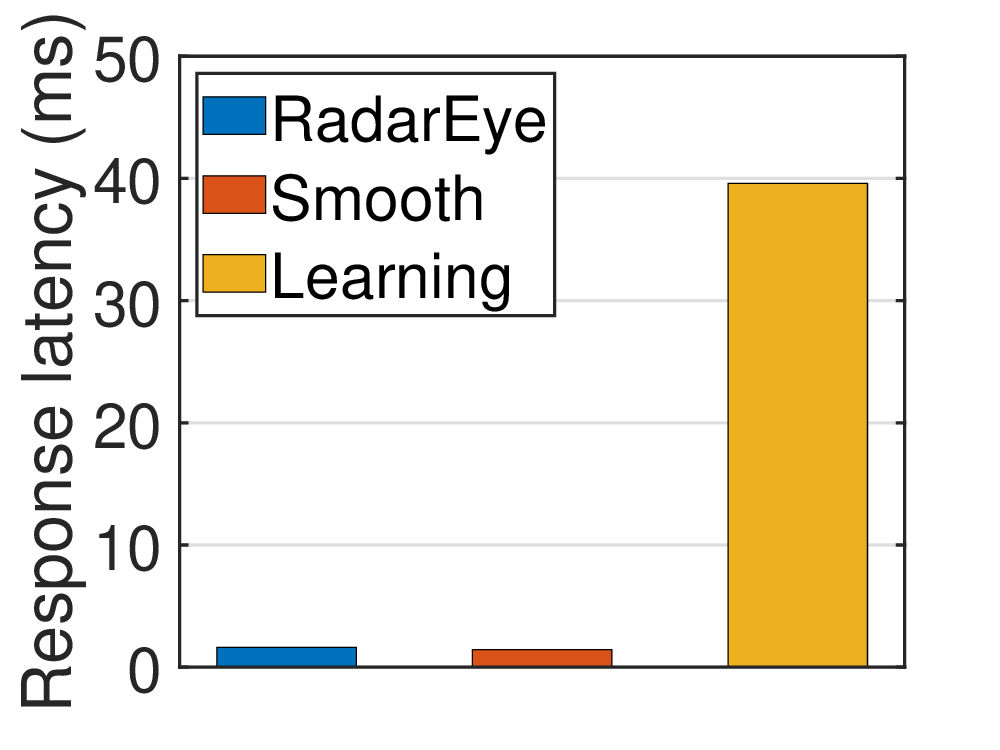} 
}
\caption{Performance comparison between RadarEye and baselines.}
\label{fig_pouring_curve}
\end{minipage}
\vspace{-1em}
\end{figure*} 

\vspace{-1em}
\section{Evaluation} \label{sec_evaluation}
\textbf{Experimental Setup}: RadarEye integrates a TI IWR6843 mmWave radar operating at a center frequency of 61.8 GHz with a 3.6 GHz bandwidth, and employs a 1$\times$4 Tx-Rx array to minimize power consumption. The radar is mounted alongside an Intel Realsense D435i RGB-D camera on a custom 3D-printed bracket, as shown in Fig.~\ref{experiment}. The manipulation tasks are performed by a Reachy humanoid robot equipped with a 7-DoF arm, operating under Ubuntu 22.04. Signal processing and robotic arm control are both implemented in Python for seamless integration within a unified software framework. Ground-truth liquid levels were obtained by monitoring a transparent ruler affixed to the inner wall of the target container using synchronized video recordings.

In our experiments, {we evaluated RadarEye’s performance in tracking water levels during pouring.}

\textbf{{Evaluating Liquid Level Estimation with Incremental Filling}}:
{To quantify the accuracy of the RadarEye system for liquid level estimation, we conducted an experiment by incrementally filling a target container with water.} Following each water addition, once the liquid surface stabilized, RadarEye was employed to extract the surface-reflected signal and estimate the liquid level. As shown in Fig.~\ref{radar_performance}, when the liquid level increased stepwise from 0 to 7.4 cm, the corresponding reflected peaks in the AoA-ToF spectrum tracked the changes accurately, with a median error of 0.12 cm. This result confirms RadarEye's ability to effectively capture surface and provide accurate liquid level estimates.

\textbf{Dynamic Liquid Level Tracking During Continues Pouring}:
We then evaluate the accuracy of the RadarEye in dynamic continues pouring tasks. We compare the performance of the RadarEye against two baseline approaches: a vision-based method and an ultrasound-based method. The experiments were conducted in a controlled setting, using  identical containers and consistent environmental conditions, to exclusively evaluate the performance differences among a radar, an RGB-D camera, and an ultrasonic transducer. The overall results are presented in Fig.~\ref{fig_pouring_curve}. Among this, Fig.~\ref{fig_pouring_curve}~(a) shows that the pouring curve generated by the RadarEye is smooth and continuous, whereas the curves from the camera and the ultrasonic sensor exhibit considerable volatility. This phenomenon occurs because, for transparent objects, the depth information from the camera is prone to missing values, which critically impairs the accuracy of liquid level estimation. The pouring curve from the ultrasonic sensor is nearly a flat line, which is attributable to the fact that acoustic sensors are typically designed for applications such as underwater communication. In contrast to mmWave signals, ultrasonic signals possess greater penetration power, causing them to detect the reflection from the bottom of the target container instead of the liquid level. Fig.~\ref{fig_pouring_curve}~(b) illustrates the water level estimation errors, and the RadarEye achieves a lowest median error of 0.35 cm, representing a significant improvement over the two baseline methods, which recorded median errors of 2.1 cm (vision-based) and 4.3 cm (ultrasound-based).

\textbf{Comparison of Level Tracking Algorithms.} 
To demonstrate the efficacy of our physics-informed algorithm for tracking liquid levels from 2D radar AoA-ToF spectrum, we conducted a comparative analysis against two baseline methods: an envelope smoothing algorithm and a compact deep learning network. The smoothing algorithm operates by averaging the top five peak reflection signals across consecutive time frames. The deep learning approach, adapted from the model proposed in Sound-Water \cite{bagad2025sound}, learns a direct mapping from the radar spectrum to the liquid level height using a collected dataset of radar-level pairs. We collect data, train the network, and evaluate performance using the same containers. Fig.~\ref{fig_pouring_curve}~(c) compares the performance of these three algorithms in response time, reveals that the deep learning algorithm exhibits significant latency, with an average response time of approximately 40 ms. In contrast, our algorithm and the smoothing algorithm are computationally efficient and achieve significantly faster response times, both approximately 0.6 ms. While response times are contingent on factors such as code optimization and the hardware environment, the inherent computational overhead of deep learning methods often makes them less suitable for time-critical tasks. Given the stringent real-time requirements of robotic pouring, the latency of the learning-based approach can severely impede task efficiency. Overall, RadarEye achieves an optimal balance, delivering superior performance in both accuracy and response time during dynamic pouring processes.

\section{Conclusion}
Transparent liquid manipulation in robotic pouring remains challenging for traditional vision-based perception systems. In this work, we propose RadarEye, a real-time mmWave radar signal processing pipeline for robust liquid height estimation and tracking during pouring process. RadarEye integrates a high-resolution 2D beamforming module for precise level estimation and a physics-informed mid-pour tracker that suppresses multipath to maintain lock on the liquid surface despite stream-induced clutter and container reflections. The pipeline delivers sub-millisecond latency. In real-robot water-pouring experiments, RadarEye achieves a 0.35 cm median error at 0.62 ms per update, substantially outperforming vision and ultrasound baselines. {Future work includes evaluating level-tracking accuracy across more transparent liquids and integrating a vision-language model to guide the robot for more robust, precise manipulation.
}

\section{Acknowledgment}
The work of He Chen is supported in part by the CUHK Strategic Seed Funding for Collaborative Research Scheme under Project 3136053 and the CUHK Direct Grant for Research under Project 4055229. The research work described in this paper was conducted in the JC STEM Lab of Advanced Wireless Networks for Mission-Critical Automation and Intelligence funded by The Hong Kong Jockey Club Charities Trust.

\bibliographystyle{IEEEbib}
\bibliography{ref}

\end{document}